\documentclass[useAMS,usenatbib,fleqn]{mn2e}
\usepackage{epsfig}
\usepackage{times}
\usepackage{amssymb}
\usepackage{multirow}
\usepackage{mathtools}
\usepackage{booktabs}

\title[J2215$+$5135 eclipses detected in LOFAR images]{Low-radio-frequency eclipses of the redback pulsar J2215$+$5135 observed in the image plane with LOFAR}
\author[J. W. Broderick et al.]{J. W. Broderick,$^{1,2,3}$\thanks{E-mail: broderick@astron.nl} R. P. Fender,$^{1,2}$ R. P. Breton,$^{4,2}$ A. J. Stewart,$^{1,2}$ A. Rowlinson,$^{3,5}$ \newauthor J. D. Swinbank,$^{6,5}$ J. W. T. Hessels,$^{3,5}$ T. D. Staley,$^{1,2}$ A. J. van der Horst,$^{7}$ M. E. Bell,$^{8,9}$ \newauthor D. Carbone,$^{5}$ Y. Cendes,$^{5}$ S. Corbel,$^{10,11}$ J. Eisl\"offel,$^{12}$ H. Falcke,$^{13,3}$ \newauthor J.-M. Grie{\ss}meier,$^{14,11}$ T. E. Hassall,$^{1,2}$ P. Jonker,$^{15,13}$ M. Kramer,$^{16,4}$ M. Kuniyoshi,$^{17}$ \newauthor C. J. Law,$^{18}$ S. Markoff,$^{5}$  G. J. Molenaar,$^{5,19}$ M. Pietka,$^{1,2}$ L. H. A. Scheers,$^{20,5}$ \newauthor M. Serylak,$^{21,22}$ B. W. Stappers,$^{4}$ S. ter Veen,$^{3}$ J. van Leeuwen,$^{3,5}$ R. A. M. J. Wijers,$^{5}$ \newauthor R. Wijnands,$^{5}$ M. W. Wise$^{3,5}$ and P. Zarka$^{23,11}$  \\ 
       $^{1}$Astrophysics, Department of Physics, University of Oxford, Keble Road, Oxford OX1 3RH, UK\\
       $^{2}$Physics and Astronomy, University of Southampton, Highfield, Southampton SO17 1BJ, UK\\
       $^{3}$ASTRON, the Netherlands Institute for Radio Astronomy, Postbus 2, 7990 AA Dwingeloo, The Netherlands \\
       $^{4}$Jodrell Bank Centre for Astrophysics, School of Physics and Astronomy, The University of Manchester, Manchester M13 9PL, UK \\
       $^{5}$Anton Pannekoek Institute for Astronomy, University of Amsterdam, Science Park 904, 1098 XH Amsterdam, The Netherlands \\
       $^{6}$Department of Astrophysical Sciences, Princeton University, Princeton, NJ 08544, USA \\  
       $^{7}$Department of Physics, The George Washington University, 725 21st Street NW, Washington, DC 20052, USA \\   
       $^{8}$CSIRO Astronomy and Space Science, PO Box 76, Epping, NSW 1710, Australia \\
       $^{9}$ARC Centre of Excellence for All-sky Astrophysics (CAASTRO), The University of Sydney, NSW 2006, Australia \\
       $^{10}$Laboratoire AIM (CEA/IRFU - CNRS/INSU - Universit\'e Paris Diderot), CEA DSM/IRFU/SAp, F-91191 Gif-sur-Yvette, France \\
       $^{11}$Station de Radioastronomie de Nan\c{c}ay, Observatoire de Paris, CNRS/INSU, USR 704 - Univ. Orl\'eans, OSUC, 18330 Nan\c{c}ay, France \\
       $^{12}$Th\"uringer Landessternwarte, Sternwarte 5, D-07778 Tautenburg, Germany \\ 
       $^{13}$Department of Astrophysics/IMAPP, Radboud University Nijmegen, PO Box 9010, 6500 GL Nijmegen, The Netherlands \\
       $^{14}$LPC2E - Universit\'{e} d'Orl\'{e}ans / CNRS, 45071 Orl\'{e}ans cedex 2, France \\
       $^{15}$SRON, Netherlands Institute for Space Research, Sorbonnelaan 2, 3584 CA, Utrecht, The Netherlands \\
       $^{16}$Max-Planck-Institut f\"{u}r Radioastronomie, Auf dem H\"{u}gel 69, 53121 Bonn, Germany \\    
       $^{17}$NAOJ Chile Observatory, National Astronomical Observatory of Japan, 2-21-1 Osawa, Mitaka, Tokyo 181-8588, Japan \\      
       $^{18}$Department of Astronomy and Radio Astronomy Lab, University of California, Berkeley, CA, USA \\
       $^{19}$Department of Physics and Electronics, Rhodes University, PO Box 94, Grahamstown, 6140 South Africa \\ 
       $^{20}$Centrum Wiskunde \& Informatica, Science Park 123, 1098 XG Amsterdam, The Netherlands \\
       $^{21}$Department of Physics \& Astronomy, University of the Western Cape, Private Bag X17, Bellville 7535, South Africa \\
       $^{22}$SKA South Africa, 3rd Floor, The Park, Park Road, Pinelands, 7405, South Africa \\
       $^{23}$LESIA, Observatoire de Paris, CNRS, UPMC, Universit\'e Paris-Diderot, 5 place Jules Janssen, 92195 Meudon, France \\
}
\begin{document}

%\date{Accepted 1988 December 15. Received 1988 December 14; in original form 1988 October 11}

\pagerange{\pageref{firstpage}--\pageref{lastpage}} \pubyear{2016}

\maketitle
\begin{abstract}
The eclipses of certain types of binary millisecond pulsars (i.e. `black widows' and `redbacks') are often studied using high-time-resolution, `beamformed' radio observations. However, they may also be detected in images generated from interferometric data. As part of a larger imaging project to characterize the variable and transient sky at radio frequencies $<200$~MHz, we have blindly detected the redback system PSR~J2215$+$5135 as a variable source of interest with the Low-Frequency Array (LOFAR). Using observations with cadences of 2 weeks -- 6 months, we find preliminary evidence that the eclipse duration is frequency dependent ($\propto \nu^{-0.4}$), such that the pulsar is eclipsed for longer at lower frequencies, in broad agreement with beamformed studies of other similar sources. Furthermore, the detection of the eclipses in imaging data suggests an eclipsing medium that absorbs the pulsed emission, rather than scattering it. Our study is also a demonstration of the prospects of finding pulsars in wide-field imaging surveys with the current generation of low-frequency radio telescopes.        
\end{abstract}

\label{firstpage}

\section{Introduction}\label{introduction} 

The so-called `black widow' and `redback' pulsars have attracted considerable interest in recent years. First discovered by \citet*[][]{fruchter88}, these binary systems consist of a millisecond pulsar and a low-mass companion star ($\lesssim 0.05$ $\rm{M}_{\odot}$ for black widows, and $\sim$$0.1$--$0.4$ $\rm{M}_{\odot}$ for redbacks); the companion, irradiated by the pulsar, is ablated \citep[e.g.][]{kluzniak88,vandenheuvel88}. Whether there is an evolutionary scenario linking the two classes remains unclear \citep*[e.g.][]{chen13,benvenuto14}, although both are thought to be a link between accreting low-mass X-ray binary systems and `recycled', rotation-powered millisecond pulsars \citep[e.g.][]{archibald09,archibald15,papitto13,patruno14,bassa14,stappers14}.   

Deep radio pulsation searches towards unidentified {\it Fermi Gamma-Ray Space Telescope} sources \citep[e.g.][]{ransom11,keith11,hessels11,ray12,bhattacharyya13}, as well as wide-field pulsar surveys and associated follow-up investigations \citep[e.g.][]{stappers96,burgay06,crawford13}, have led to the total number of known black widows and redbacks in the Galactic field increasing to over twenty at present (review by \citealt[][]{roberts13}; also see \citealt[][]{abdo13} for the second {\it Fermi} pulsar catalogue). One of these new discoveries is the redback pulsar J2215$+$5135, identified by \citet[][]{hessels11} in a $350$-MHz survey with the Green Bank Telescope (GBT). It has a spin period of $2.61$ ms, a dispersion measure (DM) of $69.2$ pc cm$^{-3}$, and a $350$-MHz flux density of $\sim$$5$ mJy. The orbital period, $4.14$ h, is very similar to that of the prototypical Galactic field redback pulsar J1023$+$0038 \citep[$4.75$ h;][]{archibald09}. 

\citet[][]{breton13} discovered the non-degenerate companion in the optical; from light curve modelling, it was found to have a minimum mass of $0.213$ $\rm{M}_{\odot}$ and to be Roche lobe filling. In a similar study by \citet[][]{schroeder14}, two modelling routines yielded a companion mass of $0.345^{+0.008}_{-0.007}$ and $0.396 \pm 0.045$ $\rm{M}_{\odot}$, respectively. Furthermore, the neutron star mass was constrained to be $ >1.75$ $\rm{M}_{\odot}$, potentially making it one of the increasing number of known neutron stars with masses significantly in excess of the $1.4$ $\rm{M}_{\odot}$ Chandrasekhar limit \citep[][]{demorest10,romani12,antoniadis13,strader15}. However, a more recent study by \citet[][]{romani15} suggests a lower neutron star mass of approximately $1.6$ $\rm{M}_{\odot}$. Reported orbital inclination measurements range from approximately $50^{\circ}$ to $90^{\circ}$ \citep[][]{breton13,schroeder14,romani15}.   

Due to the presence of an intra-binary cloud or screen of material from the ablated companion, radio eclipses are observed in black widow and redback systems \citep[e.g.][]{fruchter88,archibald09}. Although usually studied with high-time-resolution, `beamformed'\footnote{We use the term `beamformed' to describe pulsar-like observations.} observational modes, the sudden change in flux density due to eclipses can also be potentially detected in images generated from interferometric observations: if the imaging cadence is sufficiently well matched to the eclipse properties, then we would expect to see a variable compact source. In this paper, we report image-plane detections of the eclipses of PSR~J2215$+$5135 with the Low-Frequency Array \citep[LOFAR;][]{vanhaarlem13}. This source was at first blindly detected in the LOFAR `Radio Sky Monitor' (RSM; we refer the reader to \citealt[][]{fender08}, \citealt[][]{vanhaarlem13} and \citealt[][]{swinbank15}); we then later used the full RSM data set and the known orbital ephemeris of the system to construct more complete light curves. We describe the LOFAR observations and data reduction in Section~\ref{observations}, and present our results in Section~\ref{results}. In Section~\ref{discussion}, we analyse the properties of the eclipses, before concluding in Section~\ref{conclusions}. 

\section[]{LOFAR observations}\label{observations}
A detailed overview of LOFAR can be found in \citet{vanhaarlem13}. Here, we provide a summary of our subset of observations from the LOFAR RSM, a project whose main goal is to characterize the low-frequency variable and transient radio sky over very large fields of view. We used both the high-band antennas (HBA; $110$--$240$~MHz) and low-band antennas (LBA; $30$--$80$~MHz). 

\begin{table}
 \centering
  \caption{RSM beam centre coordinates for the observations used in this study. The observing strategy is described in Sections~\ref{HBA observations} and \ref{LBA observations}.}
  \begin{tabular}{ccc}
  \hline
   \multicolumn{1}{c}{Observation}  &  \multicolumn{1}{c}{RA (J2000)} & \multicolumn{1}{c}{Dec. (J2000)} \\ 
  \multicolumn{1}{c}{}  &  \multicolumn{1}{c}{(h\,\,m\,\,s)} & \multicolumn{1}{c}{($^{\circ} \,\, \arcmin \,\, \arcsec$)}  \\    
 \hline
HBA Pointing 1 & & \\
Beam\,1 & $21\,\,52\,\,30.00$   &	$+50\,\,56\,\,27.7$  \\
Beam\,2 & $21\,\,52\,\,30.00$   &	$+54\,\,51\,\,32.3$ \\
Beam\,3 & $22\,\,00\,\,00.00$   &	$+52\,\,54\,\,00.0$     \\
Beam\,4 & $22\,\,07\,\,30.00$   &	$+50\,\,56\,\,27.7$   \\
Beam\,5 & $22\,\,07\,\,30.00$   &	$+54\,\,51\,\,32.3$     \\
Beam\,6 & $22\,\,15\,\,00.00$   &	$+52\,\,54\,\,00.0$     \\
\\
HBA Pointing 2 & & \\
Beam\,1 & $22\,\,22\,\,30.00$   &	$+50\,\,56\,\,27.7$ \\
Beam\,2 & $22\,\,22\,\,30.00$   &	$+54\,\,51\,\,32.3$ \\
Beam\,3 & $22\,\,30\,\,00.00$   &	$+52\,\,54\,\,00.0$ \\
Beam\,4 & $22\,\,37\,\,30.00$   &	$+50\,\,56\,\,27.7$ \\
Beam\,5 & $22\,\,37\,\,30.00$   &	$+54\,\,51\,\,32.3$ \\
Beam\,6 & $22\,\,45\,\,00.00$   &	$+52\,\,54\,\,00.0$ \\
\\
LBA Beam\,1 & $22\,\,00\,\,00.00$   &	$+52\,\,54\,\,00.0$  \\
LBA Beam\,2 & $22\,\,15\,\,00.00$   &	$+48\,\,58\,\,55.4$ \\
\hline
\end{tabular}
\label{table:lofar_observations3}
\end{table}

\subsection{HBA}\label{HBA observations} 
Seven HBA observing runs were carried out over the period 2013 February -- 2014 January (Tables~\ref{table:lofar_observations3} and \ref{table:lofar_observations}). In each run, we obtained observations of two sets of six coordinates; these beams are part of a larger hexagonal mosaicking pattern designed to tile the entire zenith region for LOFAR (Dec. $\sim$$+53^{\circ}$). The six beams for each pointing were observed simultaneously for $2 \times 15$ min, centred on transit. Each $15$-min block comprised $2$ min on the flux calibrator, Cygnus~A, followed by $11$ min on the RSM fields. The remaining $2 \times 1$ min was necessary to switch from the calibrator to the RSM fields, and vice versa. Thus, after $30$ min, $22$ min had been spent on-source; we then immediately observed the second set of six beams.  

The `HBA Dual Inner' array configuration was used for all observations, which results in a consistent primary beam size (and hence field of view) for the Dutch `core' and `remote' stations. The primary beam full width at half maximum (FWHM) at $150$~MHz is $3.80^{\circ}$. There were a total of $36$ or $37$ Dutch stations available for each of the runs; no international stations were included. Projected baseline lengths ranged from about $43$ m -- $120$ km.  

All observations had a bandwidth of $46.9$~MHz, covered by $240$ sub-bands, each with an individual bandwidth of $195.3$ kHz. We placed $4 \times 10$ sub-bands (i.e. $4$ $\times$ $1.95$~MHz) on each of the six beams, with centre frequencies of $124$, $149$, $156$ and $185$~MHz.

\subsection{LBA}\label{LBA observations} 
Three LBA observing runs were conducted over the period 2013~August~--~2014~March (Tables~\ref{table:lofar_observations3} and \ref{table:lofar_observations2}). We used two separate beams in this study, in which PSR~J2215$+$5135 is either $2.6^{\circ}$ or $2.7^{\circ}$ from the phase centre. Each observation was $1$ h in duration; the two RSM beams and the calibrator, Cygnus~A (centred in a third separate beam), were observed simultaneously. The array configuration was `LBA Inner', which maximizes the field of view (primary beam FWHM $9.77^{\circ}$ at $60$~MHz). As for the HBA observations, a total of $36$ or $37$ Dutch stations were available for each of the runs, with the projected baselines ranging from about $91$ m -- $84$ km.  

In addition, we have used the two bands where the bandpass response is most sensitive; the centre frequencies are $54$ and $60$~MHz. The bandwidth centred on each of these frequencies, per beam, was $3.3$~MHz ($17$ sub-bands, each with individual bandwidth $195.3$ kHz).

\subsection{Data pre-processing}\label{pre-processing} 
Pre-processing was carried out using standard methods. Firstly, radio-frequency interference (RFI) was removed using {\sc aoflagger} \citep*[][]{offringa10,offringa12a,offringa12b}. Secondly, we used the `demixing' algorithm \citep*[][]{vandertol07} to subtract the response due to very bright `A-team' sources: Cassiopeia A in the high band, and both Cassiopeia~A and Cygnus~A in the low band (apart from Cygnus~A in the calibrator scans). 

Our observations were conducted during the early period of full LOFAR science operations, and, as such, optimal demixing settings had not yet been established. In the high band, the runs from 2013 March 24 -- July 14 inclusive were processed slightly differently, which afterwards was found to have resulted in poorer-quality images. For the final HBA run on 2014 January 15, we reverted to the demixing settings used in the first two runs on 2013 February 10 and March 10. 

For practical reasons concerning data volume and the computing time required for calibration and imaging, we also averaged the data in both time and frequency, typically to final resolutions of $10$~s per time-step and $4$~channels per sub-band. Bandwidth and time smearing effects \citep[e.g.][]{bridle99} are both negligible for this study.

\subsection{Calibration and imaging}\label{calibration and imaging} 

Calibration and imaging were also carried out using standard practices. The Cygnus~A data were calibrated using a preliminary low-frequency LOFAR model of the source (\citealt{mckean11}; McKean, priv. comm.). Phase-only calibration of the target fields used data from the global sky model developed by \citet[][]{scheers11}. The $74$~MHz Very Large Array (VLA) Low-Frequency Sky Survey \citep[VLSS;][]{cohen07} was the basis for these models; spectral indices\footnote{In this paper, we use the convention $S_{\nu} \propto \nu^{\alpha}$, where $S_{\nu}$ is the flux density at frequency $\nu$, and $\alpha$ is the spectral index.} were obtained by cross-correlating the VLSS with the $325$~MHz Westerbork Synthesis Radio Telescope (WSRT) Northern Sky Survey \citep[WENSS;][]{rengelink97} and the $1.4$ GHz NRAO VLA Sky Survey \citep[NVSS;][]{condon98}.  
 
For each of the separate bands and beams, primary-beam-corrected images were made with the {\sc awimager} \citep[][]{tasse13}. We used a robust weighting parameter \citep[][]{briggs95} of $0$. For the HBA data, we restricted the projected baseline range to $0.1$--$6$ k$\lambda$ ($200$ m -- $12$ km at $150$~MHz), and $0.04$--$1.2$ k$\lambda$ for the LBA observations ($200$ m -- $6$ km at $60$~MHz). These baseline ranges were determined empirically to ensure reliable images, given both the limitations of the $uv$ coverage per run, as well as our relatively rudimentary calibration procedure.   

We concatenated the $2 \times 11$ min HBA snapshots prior to imaging; the resulting maps for the six beams were then mosaicked together to create the final set of images, i.e. separate $22$-min mosaics at each of the four central frequencies, per pointing. In the low band, we divided the $60$-min data sets into $3 \times 20$ min segments, and created separate maps for each frequency and beam. The median angular resolutions are approximately $60$ arcsec $\times$ $30$ arcsec (HBA), and $280$ arcsec $\times$ $140$ arcsec (LBA). 

In the 2013 February 10, 2013 March 10 and 2014 January 15 HBA mosaics, the typical rms noise level near PSR~J2215$+$5135 is about $10$--$15$ mJy beam$^{-1}$. However, on average, it is a factor of $\sim$$2$--$2.5$ worse in the remaining HBA runs, where the demixing settings had been changed. In the LBA, the noise level ranges from about $190$--$660$ mJy beam$^{-1}$. For both the HBA and LBA observations, the noise levels are above the nominal confusion limits by factors of $\sim$$2$--$4$ in the best-case scenarios. This is likely to be a consequence of both our simple calibration strategy and residual signal from Cassiopeia~A, which is only $12^{\circ}$ from PSR~J2215$+$5135. Attempts at further phase-only self-calibration did not result in significant differences to either the image dynamic range or the source flux densities.    
  
We estimate that the internal calibration uncertainty across all of the runs ranges from $15$--$20$ per cent in the high band, depending on the demixing settings used. In the low band, it is about $30$ per cent. The absolute flux scale is that of \citet[][]{baars77}.

\begin{table*}
 \centering
 \begin{minipage}{160mm}
  \caption{LOFAR high-band image-plane observations of PSR~J2215$+$5135. MJD is the Modified Julian Date at the halfway point of each observation (in Coordinated Universal Time; \textsc{utc}), and $\phi$ is the orbital phase determined using the ephemeris of \citet[][]{abdo13} and Hessels et al. (in prep.). Orbital phases are defined such that the superior conjunction of the pulsar is at $\phi=0.5$. A description of how the flux densities were measured can be found in Section~\ref{results}. Spectral indices across the high band were determined from linear least-squares fits in $\log(S_{\nu})$--$\log(\nu)$ space, with inverse-variance weighting. All uncertainties are $\pm 1\sigma$. We also give the LOFAR observation IDs for each epoch.}
  \begin{tabular}{cccrrrrrl}
  \hline
   \multicolumn{1}{c}{Date}  &  \multicolumn{1}{c}{MJD} & \multicolumn{1}{c}{$\phi$} & \multicolumn{1}{c}{$S_{124 \, \rm MHz}$} &  \multicolumn{1}{c}{$S_{149 \, \rm MHz}$} & \multicolumn{1}{c}{$S_{156 \, \rm MHz}$} &  \multicolumn{1}{c}{$S_{185 \, \rm MHz}$} & \multicolumn{1}{c}{$\alpha_{\rm HBA}$} & \multicolumn{1}{c}{IDs} \\
    &  & (range)  & \multicolumn{1}{c}{(mJy)} & \multicolumn{1}{c}{(mJy)} & \multicolumn{1}{c}{(mJy)} & \multicolumn{1}{c}{(mJy)} & & \multicolumn{1}{c}{(range)} \\ 
 \hline
2013 February 10 & $56333.509$ & $0.381$--$0.486$ & $18 \pm 15$ 	& $-15 \pm 9$	& $1 \pm 9$	& $-4 \pm 9$ & & L$89610$-- \\ 
 & $56333.530$ & $0.502$--$0.606$ & $3 \pm 13$ & $-1 \pm 10$ &	$5 \pm 10$ & $8 \pm 12$ & & L$89617$ \\
\\
2013 March 10 & $56361.431$ & $0.242$--$0.347$ & $57 \pm 20$ &	$58 \pm	14$ & $47 \pm 13$ & $48 \pm 12$ & $-0.5 \pm 1.0$ & L$100346$-- \\ 
& $56361.452$ & $0.363$--$0.467$ & $7 \pm 16$ & $3 \pm 11$	& $6 \pm 12$	& $3 \pm 12$ & & L$100353$ \\
\\
2013 March 24 & $56375.392$ & $0.174$--$0.278$ & $168 \pm 51$ &	$131 \pm 33$ & $122 \pm 32$ & $103 \pm 30$ & $-1.2 \pm 1.0$ & L$107837$-- \\  
& $56375.413$ & $0.294$--$0.399$ & $1 \pm 21$	& $8 \pm 18$	& $13 \pm 21$	& $-70 \pm 51$ & & L$107844$	\\
\\
2013 April 22 & $56404.314$ & $0.837$--$0.941$ & $202 \pm 42$ &	$123 \pm 38$ & $108 \pm 34$ & $104 \pm 30$ & $-1.8 \pm 0.9$ & L$124005$--  \\ 
& $56404.335$ & $0.958$--$0.062$ & $208 \pm 69$ & $176 \pm 43$ & $119 \pm 43$ & $91 \pm 45$ & $-2.1 \pm 1.4$  & L$124012$ \\ 
\\
2013 May 20 & $56432.238$ & $0.715$--$0.820$ & $171 \pm	37$ & $116 \pm 40$ & $93 \pm 35$	& $99 \pm 31$	& $-1.5 \pm 0.9$ & L$133624$--   \\ 
& $56432.258$ & $0.836$--$0.940$ & $256 \pm 75$ & $161 \pm 35$ &	$141 \pm 37$ & 	$73 \pm 47$ & $-2.8 \pm 1.4$ & L$133643$\rlap{$^{\rm a}$} \\
\\
2013 July 14 & $56487.087$ & $0.695$--$0.800$ & $116 \pm 36$ & $68 \pm 36$ & $38 \pm 32$ & $39 \pm 32$ & & L$160506$--  \\ 
& $56487.108$ & $0.816$--$0.921$ & $183 \pm 41$ & $164 \pm 57$ & $82 \pm 50$	& $91 \pm 47$ & & L$160513$ \\
\\
2014 January 15 & $56672.578$ & $0.984$--$0.089$ & $228 \pm 37$ & $129 \pm 22$ & $132 \pm 22$ & $89 \pm	16$ & $-2.3 \pm 0.6$ & L$198919$--  \\ 
& $56672.599$ & $0.104$--$0.209$ & $237 \pm 39$ & $148 \pm 25$ &	$126 \pm 23$ & $86 \pm 18$ & $-2.6 \pm 0.6$ & L$198926$ \\
\hline
\multicolumn{9}{l}{{$^{\rm a}$} The relevant IDs are L$133624$--L$133625$, L$133630$--L$133631$, L$133637$--L$133638$ and L$133642$--L$133643$.}   
\end{tabular}
\label{table:lofar_observations}
\end{minipage}
\end{table*}  

\begin{table*}
\begin{minipage}{160mm}
 \centering
  \caption{LOFAR low-band image-plane observations of PSR~J2215$+$5135. The flux densities are the averages from the two beams used in this study (Table~\ref{table:lofar_observations3}). Otherwise, see Table~\ref{table:lofar_observations} for a description of the columns.}
  \begin{tabular}{cccrrc}
  \hline
   \multicolumn{1}{c}{Date}  &  \multicolumn{1}{c}{MJD} & \multicolumn{1}{c}{$\phi$} & \multicolumn{1}{c}{$S_{54 \, \rm MHz}$} & \multicolumn{1}{c}{$S_{60 \, \rm MHz}$} & \multicolumn{1}{c}{ID} \\
   &  & (range) & \multicolumn{1}{c}{(mJy)} & \multicolumn{1}{c}{(mJy)} & \\ 
 \hline
2013  & $56515.995$ & $0.294$--$0.373$ & $-120 \pm 170$ & $40 \pm 140$ & L$167732$ \\  
August 11/12& $56516.008$ & $0.373$--$0.452$ & $-30 \pm 180$ & $530 \pm 220$ &	 \\  
& $56516.022$ & $0.452$--$0.532$ & $-290 \pm 170$ & $-320 \pm 200$ &	 \\  
\\
2013  & $56594.778$ & $0.009$--$0.088$ & $1570 \pm 510$ & $1980 \pm 680$ & L$183339$  \\ 
October 29& $56594.792$ & $0.088$--$0.167$ & $2980 \pm 920$ & $2090 \pm 710$ &  \\
& $56594.806$ & $0.167$--$0.246$ & $2680 \pm 830$ & $1230 \pm 560$ &  \\ 
\\
2014  & $56746.365$ & $0.726$--$0.805$ & $-190 \pm 270$ & $100 \pm 180$ & L$214818$ \\ 
March 30 & $56746.379$ & $0.805$--$0.885$ & $1980 \pm 640$ & $1650 \pm 530$ & \\
& $56746.392$ & $0.885$--$0.964$ & $2700 \pm 840$ & $1890 \pm 590$ & \\
\hline
\end{tabular}
\label{table:lofar_observations2}
\end{minipage}
\end{table*}

\section{Results}\label{results}

\begin{figure*}
\begin{minipage}{180mm}
\epsfig{file=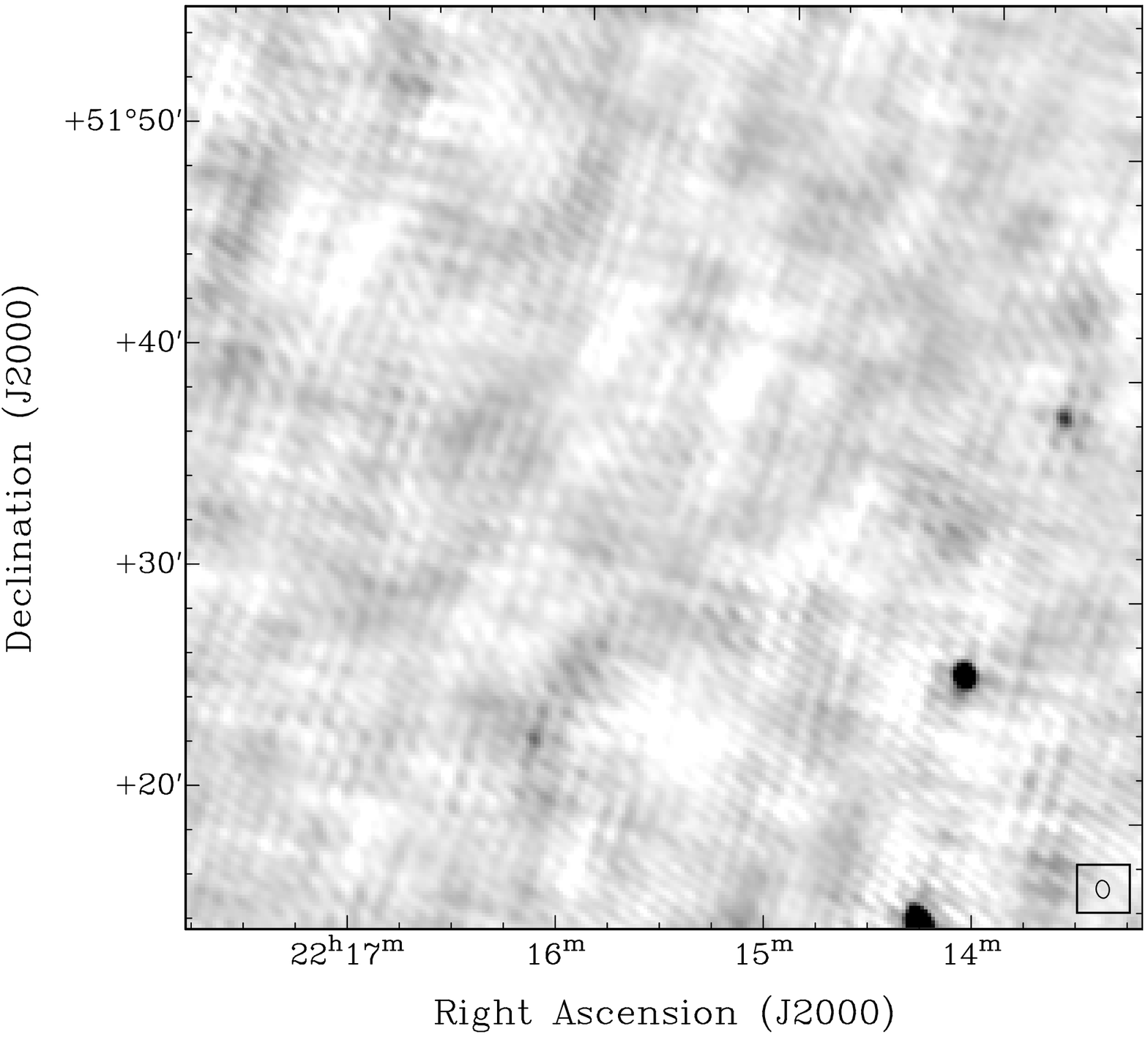,width=8cm}
\hspace{1cm}
\epsfig{file=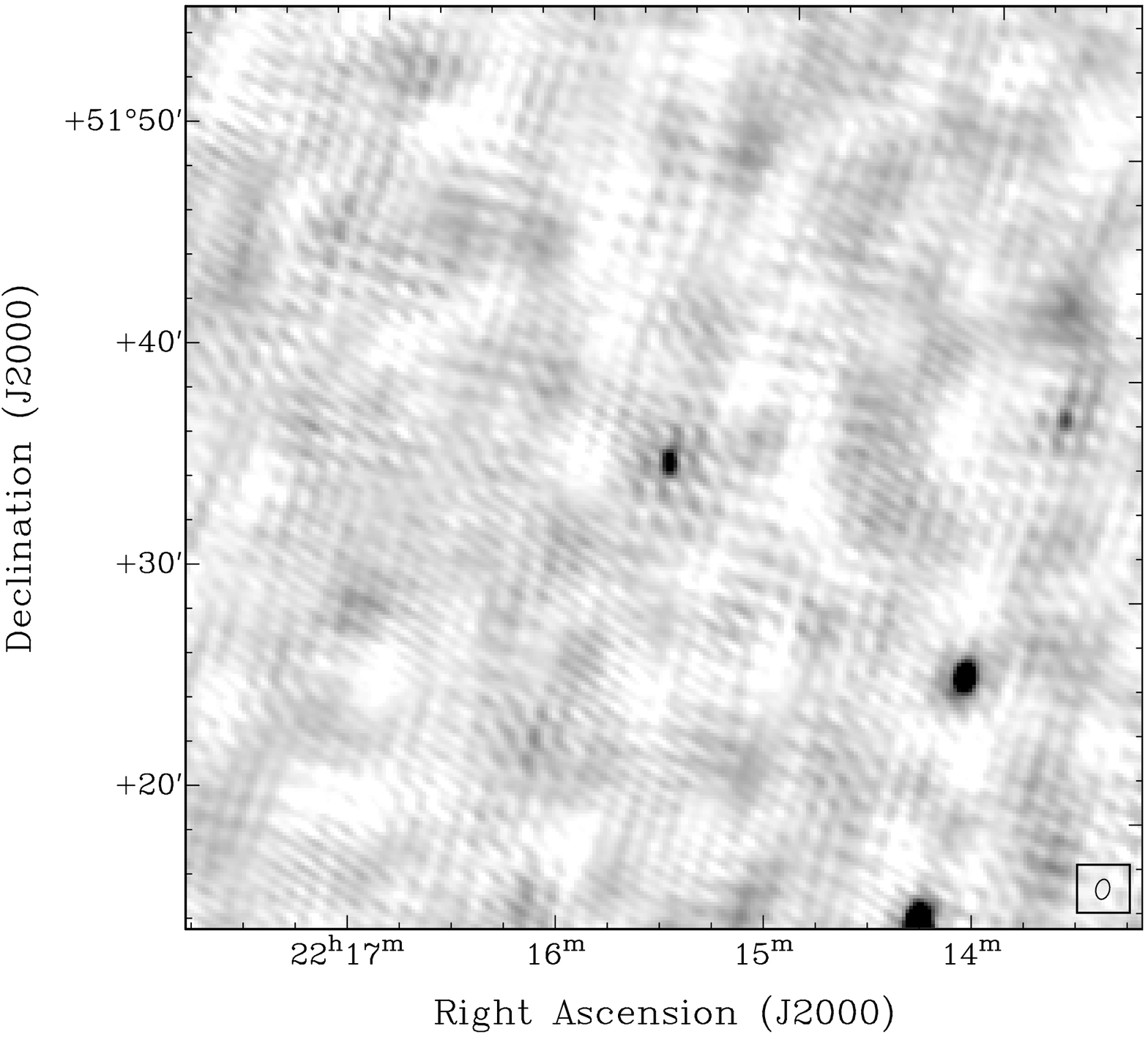,width=8cm}
\caption{{\em Left}: $124$-MHz LOFAR map from 2013 February 10 (MJD $56333.509$), zoomed in on the position of PSR~J2215$+$5135. The angular resolution is $48$ arcsec $\times$ $35$ arcsec (synthesized beam position angle $10^{\circ}$), and the rms noise level is $13$ mJy beam$^{-1}$. The pulsar is not detected. {\em Right}: The same field, but from the observation on 2014 January 15 (MJD $56672.578$). In this case, the angular resolution is $55$ arcsec $\times$ $37$ arcsec (beam position angle $-11^{\circ}$), and the rms noise level is $15$ mJy beam$^{-1}$. PSR~J2215$+$5135 is now detected at approximately $15$ times the local rms noise level, with a flux density of $228$ mJy. The synthesized beam is shown in the bottom right-hand corner of each panel.}
\label{fig:pulsar1}
\end{minipage}
\end{figure*}

\begin{figure*}
\begin{minipage}{180mm}
\centering
\epsfig{file=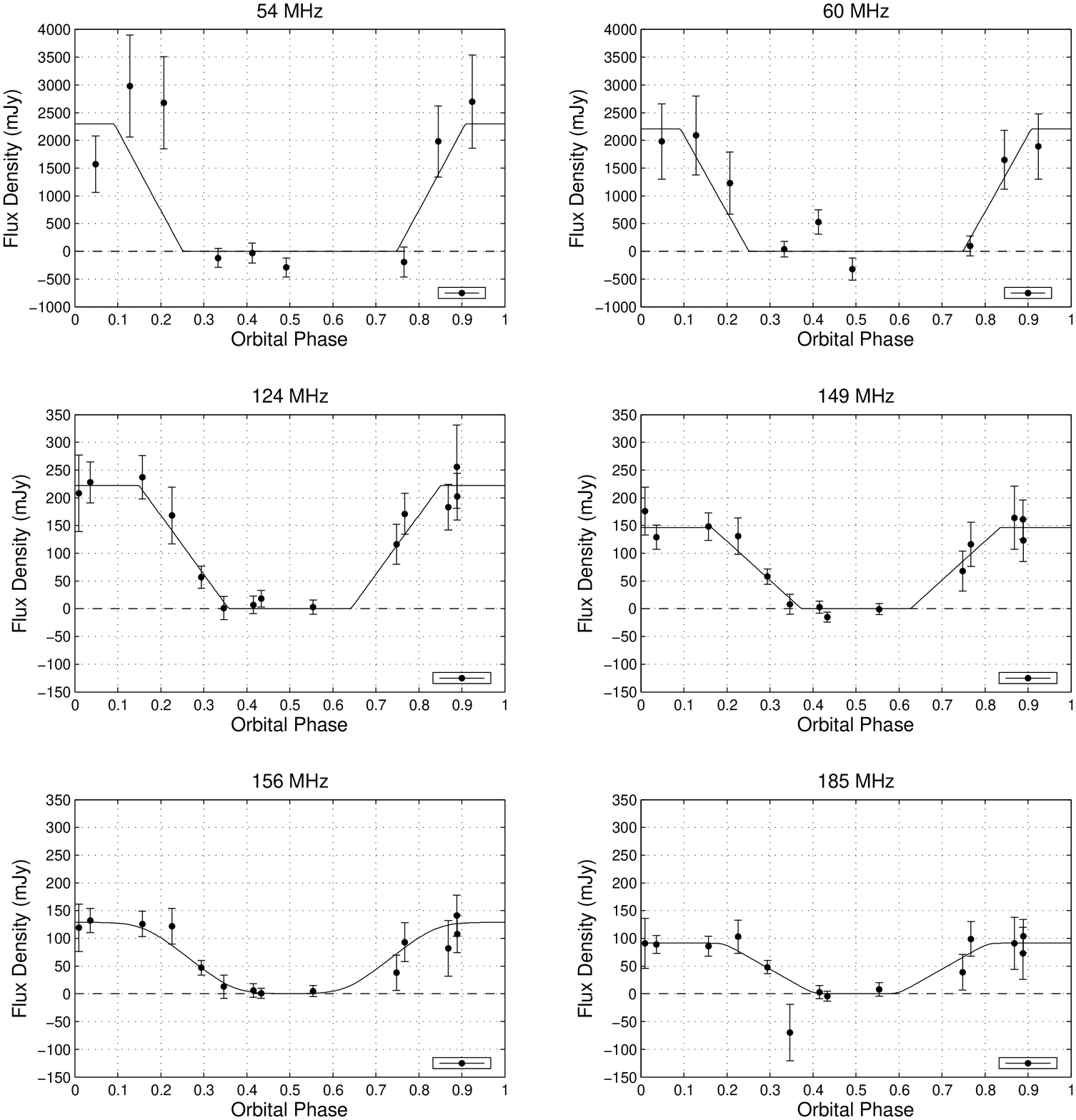,width=16cm}
\caption{Flux density as a function of orbital phase (as defined in Equation~\ref{eqn:phase}, with the pulsar superior conjunction at $\phi=0.5$) for the six frequencies used in this study. The average orbital phase is plotted for each separate observation; in the bottom right-hand corner of each plot, we indicate the phase range per data point. For the HBA data sets, the integration time on-source per data point is $22$ min, although note that a $26$-min scan duration is used for calculating the phase range, given that there is a calibrator observation between the two, $11$-min snapshots (see Section~\ref{HBA observations} for further details). For the LBA panels, the integration time per data point is $20$ min. In each panel, the solid line is a fitted double Fermi-Dirac-like function using the set of parameters for which the posterior probability was maximized (Section~\ref{eclipse properties} and Table~\ref{table:lofar_observations5}). Each model profile also takes into account that the eclipse is smeared out due to the limited sampling; this means that a very sharp eclipse becomes `ramp-like', a shape that is seen in the majority of the profiles. On the other hand, smoother underlying profiles (e.g. at 156 MHz) are smoothed further by this effect; see Section~\ref{eclipse properties} for futher details. All flux density error bars are $\pm1\sigma$, while the dashed line indicates a zero-level flux density.}
\label{fig:pulsar2}
\end{minipage}
\end{figure*}

We detected a variable source at the following coordinates: RA\,\,$22^{\rm h}15^{\rm m}32\fs4$, Dec.\,\,$+51^{\circ}35\arcmin39\arcsec$ (J2000). We estimate that the position is accurate to within approximately $5$ arcsec (in the high band); thus, the coordinates are in agreement with the position of PSR~J2215$+$5135 derived from beamformed observations (\citealt[][]{abdo13}; Hessels et al. in prep.). 

Initially, the discovery was made from visual inspection of the HBA data from 2013 February 10 and March 10, as well as 2014 January 15 (Figure~\ref{fig:pulsar1}). The variability was subsequently confirmed with the {\sc transients pipeline} \citep[{\sc trap};][]{swinbank15}. Then, as mentioned in Section~\ref{introduction}, we used the full RSM data set to obtain more complete source properties for PSR~J2215$+$5135.  
  
Measured flux densities from point-source Gaussian fits are presented in Tables~\ref{table:lofar_observations}~and~\ref{table:lofar_observations2}. We used {\sc imfit} in {\sc miriad} \citep*[][]{sault95} for this task, and confirmed the results using {\sc pyse} \citep[][]{spreeuw10}. For the low signal-to-noise (S/N) detections, we report the fitted flux density from a forced point-source fit at the source location, rather than an upper limit, to allow for improved light curve modelling (Section~\ref{eclipse properties}). In a handful of the noisiest images, we instead estimated the flux density based on the integrated pixel sum in a box surrounding the source position. All flux densities have been corrected for the background level in the vicinity of the source, and the uncertainty in this process is included in the overall flux density uncertainty, along with the calibration uncertainty and the statistical fitting error (all terms added in quadrature). Biases arising from the different techniques used are an additional, but second-order effect. Note that noise fluctuations and/or uncertainties in the background level can occasionally result in negative flux densities for the forced fits; however, all such cases in this paper are consistent with a zero-level flux density within $1.7\sigma$ at most. 

On 2013 March 24, there were significant differences between the high-band flux densities in the two pointings (spaced by $30$ min). Similar behaviour was seen on 2014 March 30 in the low band. The modulation index, $\sigma_{S}/\overline{S}$, where $\sigma_{S}$ is the standard deviation of the flux densities and $\overline{S}$ is the mean flux density, ranges from about $0.7$--$1.1$ for the full set of observations at each of the six different frequencies.  

In Table~\ref{table:lofar_observations}, we have also calculated the in-band spectral index for each HBA run where there were at least three detections $> 3\sigma$ (relative to the local rms noise level, and not including the additional uncertainties discussed above). Typically, the source has a very steep spectrum between $124$ and $185$~MHz, although the uncertainties for the individual measurements are often significant. The median spectral index is $\sim$$-2$.

\section{Discussion}\label{discussion}
    
\subsection{Evidence for eclipses}\label{eclipse evidence} 

We now show that a pulsar origin is consistent with the source properties obtained from the LOFAR image-plane data. Firstly, the orbital phase, $\phi$, of PSR~J2215$+$5135 can be determined using the equation
\begin{equation}\label{eqn:phase}
\phi = \left[\frac{(t_{\rm obs} - T_{\rm asc})\bmod{P_{\rm orb}}}{P_{\rm orb}} + 0.25 \right]\bmod{1}. 
\end{equation} 
In Equation~\ref{eqn:phase}, $t_{\rm obs}$ is the Modified Julian Date (MJD) of the observation, with an additional barycentric correction; we used the software tools from \citet*[][]{eastman10} for this task. The orbital period $P_{\rm orb} = 0.172502104907(8)$ d, where the number in parentheses indicates the uncertainty on the least significant digit, and the time of the ascending node $T_{\rm asc} = 55186.164486(1)$ (\citealt[][]{abdo13}; Hessels et al. in prep.). In our definition of orbital phase, the superior conjunction of the pulsar occurs at $\phi=0.5$. 

We calculated $\phi$ for each observation using Equation~\ref{eqn:phase}; these values are given in Tables~\ref{table:lofar_observations} and \ref{table:lofar_observations2}. Although the individual data points sample $\sim$$10$ per cent of the orbit only, our combined HBA and LBA data sets provide good coverage across the full range of orbital phases. 

Figure~\ref{fig:pulsar2} shows flux density versus orbital phase for each of the six different frequencies. In each case, the observed change in the flux density implies that the variability is due to the eclipses of PSR~J2215$+$5135.

\subsection{Modelling the eclipses}\label{eclipse properties}

\begin{table*}
\begin{minipage}{180mm}
 \centering 
  \caption{Properties of the parameter distributions used in the fitting of the light curves in Figure~\ref{fig:pulsar2}. For each parameter and frequency, we first list the value from the set that yielded the maximum posterior probability (MPP). This is followed by the median and mean, including the formal $1\sigma$ uncertainty for the latter.}
  \begin{tabular}{crrrrrrrrrrrr}
  \hline
\multicolumn{1}{c}{Frequency}  & & \multicolumn{3}{c}{$S_{0}$ (mJy)}  & & \multicolumn{3}{c}{$\mu$}  & & \multicolumn{3}{c}{$\beta$}  \\\cmidrule{3-5}\cmidrule{7-9}\cmidrule{11-13}   
\multicolumn{1}{c}{(MHz)}  & & \multicolumn{1}{c}{MPP} & \multicolumn{1}{c}{Median} & \multicolumn{1}{c}{Mean} & & \multicolumn{1}{c}{MPP} & \multicolumn{1}{c}{Median} & \multicolumn{1}{c}{Mean} & & \multicolumn{1}{c}{MPP} & \multicolumn{1}{c}{Median} & \multicolumn{1}{c}{Mean}  \\ 
 \hline
$54$   & & $2300$ & $2380$ & $2600 \pm 990$   & & $0.172$ & $0.165$ & $0.152 \pm 0.044$   & & $6.6 \times 10^{-4}$ & $0.019$ & $0.025 \pm 0.020$  \\ 
$60$    & & $2210$ & $2870$ & $3410 \pm 1640$   & & $0.172$ & $0.127$ & $0.114 \pm 0.057$   & & $9.2 \times 10^{-5}$ & $0.041$ & $0.042 \pm 0.026$  \\   
$124$   & & $223$ & $234$ & $240 \pm 40$   & & $0.254$ & $0.244$ & $0.240 \pm 0.027$   & & $3.4 \times 10^{-4}$ & $0.024$ & $0.027 \pm 0.019$       \\ 
$149$   & & $147$ & $149$ & $149 \pm 14$   & & $0.269$ & $0.266$ & $0.266 \pm 0.018$   & & $4.4 \times 10^{-4}$ & $0.014$ & $0.016 \pm 0.012$  \\
$156$   & & $129$ & $135$ & $141 \pm 36$   & & $0.264$ & $0.253$ & $0.246 \pm 0.042$   & & $0.027$ & $0.035$ & $0.037 \pm 0.024$  \\
$185$   & & $91$ & $95$ & $96 \pm 15$   & & $0.297$ & $0.286$ & $0.283 \pm 0.034$   & & $8.5 \times 10^{-3}$ & $0.023$ & $0.027 \pm 0.021$ \\ 
\hline
\end{tabular}
\label{table:lofar_observations5}
\end{minipage}
\end{table*}

We quantified the eclipse properties by using a Markov Chain Monte Carlo (MCMC) approach, fitting a double Fermi-Dirac-like function of the form
\begin{equation}
S(\phi) = S_{0} \left [ \frac{1}{e^{(\phi - \mu_{1})/\beta_{1}} + 1} + \frac{1}{e^{-(\phi - \mu_{2})/\beta_{2}} + 1} \right] + S_{\rm b},   
\label{light curve fits} 
\end{equation}
to each light curve in Figure~\ref{fig:pulsar2}. In Equation~\ref{light curve fits}, $S(\phi)$ is the flux density at orbital phase $\phi$, $S_{0}$ is the uneclipsed, pulsed flux density, and $S_{\rm b}$ is the unpulsed, `baseline' flux density. The parameters $\mu_{1}$ and $\mu_{2}$ are the orbital phases at which the pulsed flux density is half of its maximum value (i.e. $S_{0}/2$) during eclipse ingress and egress, respectively. Moreover, the constants $\beta_{1}$ and $\beta_{2}$ describe the slopes of ingress and egress, respectively; a smaller value corresponds to a more rapid, sharper transition. 

We also accounted for the smearing of the eclipse profile due the integration time per data point being $\sim$$10$ per cent of the orbit. To do so, for each realisation of the eclipse light curve, we generated a high-resolution model containing $500$ samples across an orbit. We then convolved it with a boxcar kernel having a width equal to the integration time, and using cyclic boundaries in order to ensure continuity and smoothness. For very small values of $\beta_{1}$ and/or $\beta_{2}$, this results in a sharp eclipse becoming `ramp-like' (recall that the convolution of a Heaviside step function with a boxcar produces such a functional shape). For larger values of $\beta_{1}$ and/or $\beta_{2}$, both the underlying and smeared eclipse profiles are smoother.   

Several possible variants of Equation~\ref{light curve fits} were trialled, with the number of free parameters adjusted. To distinguish between the different possibilities, we used the Akaike Information Criterion \citep[e.g.][]{akaike74,cavanaugh97,burnham04}. We found that the data quality is not sufficient for increasingly sophisticated models. Hence, we used a simple model in which (i) there is no unpulsed component; (ii) the eclipses are symmetrical in time about the superior conjunction of the pulsar; and (iii) eclipse ingress and egress have symmetrical slopes. Equation~\ref{light curve fits} can therefore be rewritten as 
\begin{equation}
S(\phi) = S_{0} \left [ \frac{1}{e^{(\phi - \mu)/\beta} + 1} + \frac{1}{e^{-(\phi - (1 - \mu))/\beta} + 1} \right],   
\label{light curve fits2} 
\end{equation}
where we have defined $\beta_{1} = \beta_{2} \equiv \beta$, and $1-\mu_{2} = \mu_{1} \equiv \mu$.

The model fits are overplotted on the light curves in Figure~\ref{fig:pulsar2}. Each fit was generated using the set of parameters with the maximum posterior probability value; we list these in Table~\ref{table:lofar_observations5}. We also report the medians and means of the parameter distributions in Table~\ref{table:lofar_observations5}; note that the distributions are sometimes skewed and the parameters correlated with each other. In particular, $\beta$ is right skewed. For simplicity, and given the limitations of our analysis, we will use the mean values for $S_0$, $\mu$ and $\beta$ in the rest of this paper; our conclusions are not affected significantly by this choice.   

The $\beta$ values in Table~\ref{table:lofar_observations5} are generally very small. Therefore, as explained above, the models for the majority of the light curves are not smooth. Moreover, all the $\beta$ measurements are consistent with zero within the formal $95$ per cent confidence interval, a value that would imply a perfectly sharp eclipse. Finer sampling is needed to quantify the eclipse sharpness more accurately.

\subsection{Eclipse properties}\label{eclipse properties2} 

\begin{figure}
\centering
\epsfig{file=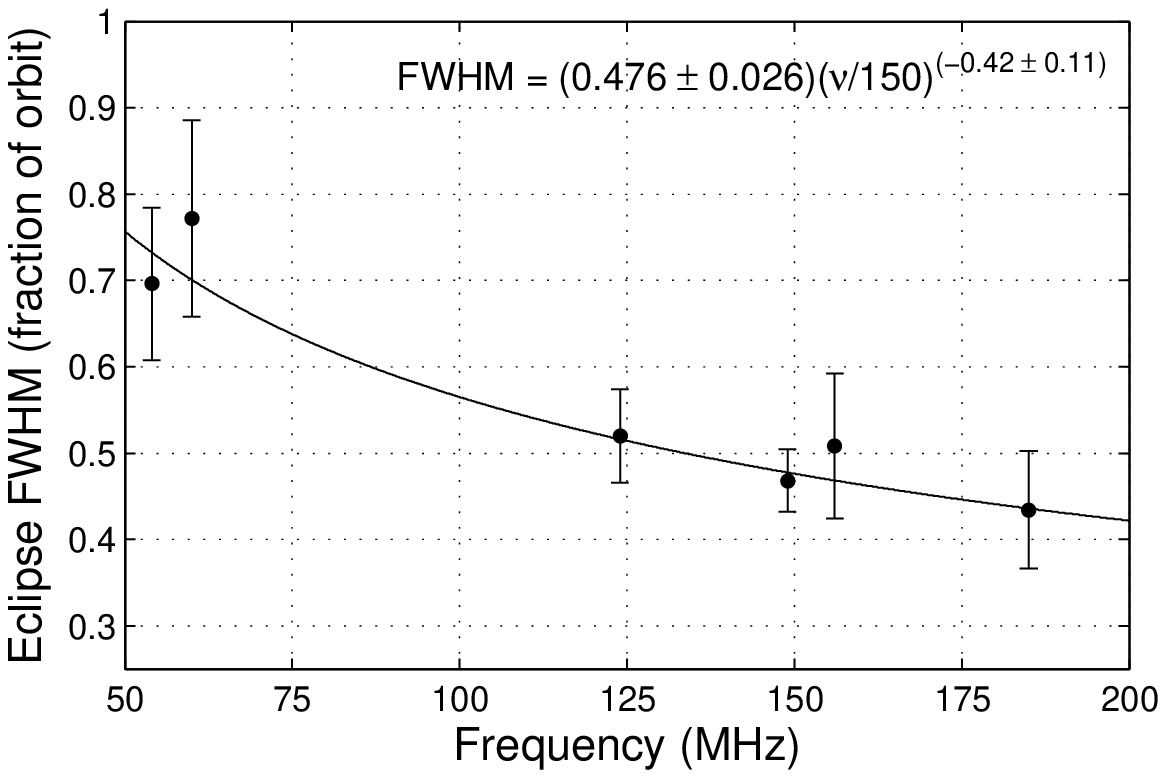,width=8cm}
\caption{Eclipse FWHM, $1-2\mu$, as a function of frequency. The FWHMs are deduced from the mean $\mu$ values in Table~\ref{table:lofar_observations5}. A power-law fit is also shown (reduced $\chi^2=0.22$).}
\label{fig:pulsar5}
\end{figure}

In Figure~\ref{fig:pulsar2}, the model light curves are eclipsed for longer at LBA frequencies. The frequency dependence is quantified in Figure~\ref{fig:pulsar5}, where we have plotted the eclipse FWHM, $1-2\mu$, as a function of frequency. Fitting a power law to the data, the anti-correlation is found to have a frequency dependence of approximately $\nu^{-0.4}$: the source is eclipsed for $\sim$$40$--$50$ per cent of the orbit ($\sim$$1.7$--$2.1$ h) in the high band, but $\sim$$70$--$80$ per cent ($\sim$$2.9$--$3.3$ h) in the low band.    

Other redback and black widow systems are known to be eclipsed for longer at lower frequencies \citep[e.g.][]{archibald09,archibald13,bhattacharyya13}. Various scenarios for an eclipsing mechanism are discussed, for example, in \citet[][]{thompson94}. Our image-plane observations suggest that the eclipses could be due to absorption, rather than scattering of the pulsations to a degree that greatly exceeds the pulse period, as the latter would still lead to image-plane detections. Indeed, \citet[][]{bhattacharyya13} and \citet[][]{roy15}, the latter using simultaneous imaging and beamformed data, proposed that cyclotron-synchrotron absorption may be responsible for the eclipses of PSR~J1544$+$4937 and PSR~J1227$-$4853~/~XSS~J12270$-$4859.

In the case of PSR~J2215$+$5135, observations that sample full individual orbits, with much finer orbital phase resolution, are needed to confirm the anti-correlation of eclipse length with frequency, which we regard as preliminary given the current evidence. This is particularly due to the relatively small number of data points, and also because the eclipses may be dynamic, with properties changing from orbit to orbit \citep[e.g.][]{archibald09,archibald13,bhattacharyya13}. 

\subsection{Radio spectrum}
\begin{figure*}
\begin{minipage}{180mm}
\centering
\epsfig{file=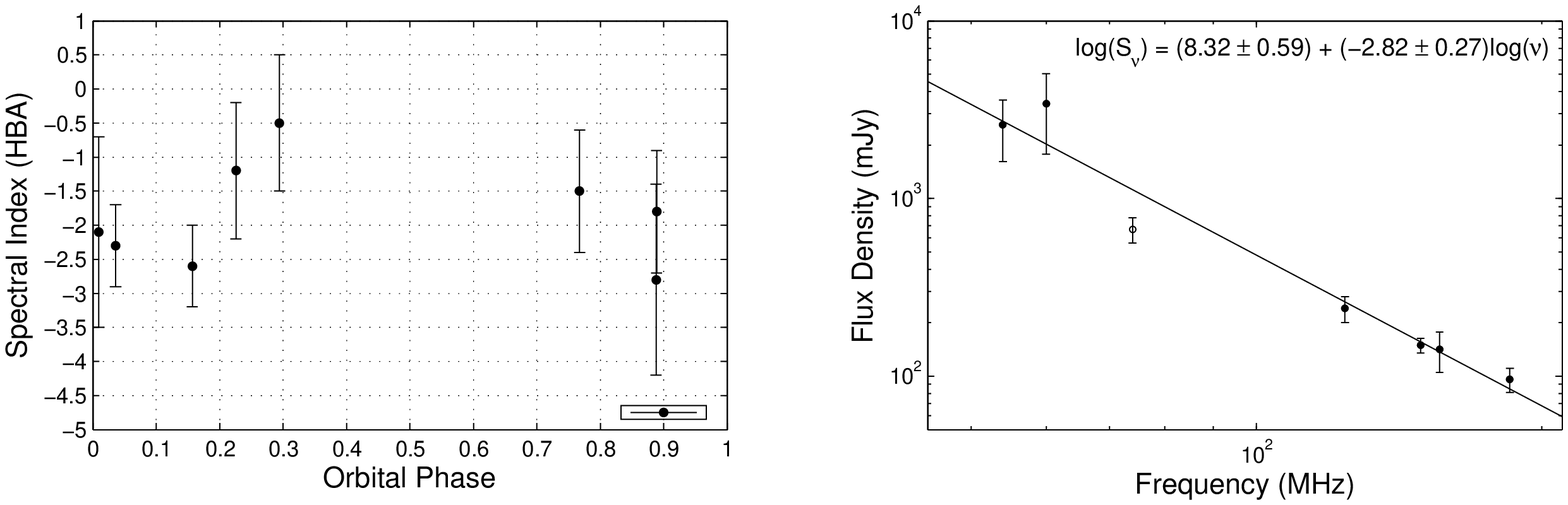,width=16cm}
\caption{{\em Left}: Spectral index in the high band, as a function of orbital phase. We follow the same conventions as in Figure~\ref{fig:pulsar2}. {\em Right:} Radio spectrum for PSR~J2215$+$5135. The filled circles are the LOFAR mean fitted flux densities from Table~\ref{table:lofar_observations5}, and the open circle is the $74$-MHz VLSSr flux density, adjusted to the scale of \citet[][]{baars77}. A linear least-squares fit in $\log(S_{\nu})$--$\log(\nu)$ space, with inverse-variance weighting, to the six LOFAR data points, is shown (solid line); the reduced $\chi^2$ value is $0.59$.}
\label{fig:pulsar4}
\end{minipage}
\end{figure*}

We plot the high-band spectral index as a function of orbital phase in the left panel of Figure~\ref{fig:pulsar4}. No statistically significant correlations are found, although there is a very tentative hint of spectral flattening during eclipse ingress, which would be consistent with the pulsar being eclipsed for longer at lower frequencies. 

A LOFAR radio spectrum from $54$--$185$ MHz, derived using the mean fitted flux densities from Table~\ref{table:lofar_observations5}, is shown in the right panel of Figure~\ref{fig:pulsar4}. The spectral index is $-2.82 \pm 0.27$. The predicted $74$-MHz flux density is about 70 per cent higher than the recently-catalogued VLSS Redux \citep[VLSSr;][]{lane14} measurement, the latter being $670 \pm 110$ mJy when placed on the \citet[][]{baars77} scale. However, the orbital phases covered by the VLSS suggest that the pulsar was partially eclipsed on average, potentially explaining the discrepancy.

\section{Conclusions}\label{conclusions}

Using interferometric observations in both the LOFAR high and low bands, we have detected the eclipses of the redback millisecond pulsar J2215$+$5135 in the image plane. By modelling the light curves at six different frequencies, we find preliminary evidence that the eclipse duration is longer at lower frequencies, with a frequency dependence of approximately $\nu^{-0.4}$. 

The fact that the eclipses are observed in our imaging data suggests that the pulsed emission is being absorbed by the eclipsing screen(s), as scattering would still lead to image-plane detections of the pulsar when it is at superior conjunction. Whether PSR~J2215$+$5135 shares a common eclipsing mechanism with other redbacks and black widows remains to be determined.  
  
Current wide-field, low-frequency, image-plane searches for variables and transients \citep[e.g.][]{bell14,obenberger14,stewart16} have the potential to be able to uncover more redbacks and black widows, as well as other types of pulsars, particularly those that are very steep spectrum (such as PSR~J2215$+$5135), highly polarized \citep[e.g.][]{navarro95}, and/or highly scattered by the intervening interstellar medium along the line of sight (e.g. \citealt{dembska15}). Simultaneous image-plane and beamformed observations \citep[e.g.][]{stappers11,roy15} will be the most robust method for studying these systems in the radio domain.

\section*{Acknowledgements}

We thank the referee for their comments and suggestions that significantly improved the presentation of this paper. We also thank the ASTRON Radio Observatory, particularly Michiel Brentjens and Carmen Toribio, for their considerable efforts in setting up the observations and pre-processing the raw data. 

This project was supported by European Research Council Advanced Grant 267697 ``4 Pi Sky: Extreme Astrophysics with Revolutionary Radio Telescopes", European Research Council Advanced Grant 247295 ``AARTFAAC'', and the European Union Seventh Framework Programme under grant agreement PIIF-GA-2012-332393. J.W.T.H. acknowledges funding from a NWO Vidi fellowship and European Research Council Starting Grant ``DRAGNET'' (337062). S.C. acknowledges financial support from the UnivEarthS Labex program of Sorbonne Paris Cit\'e (ANR-10-LABX-0023 and ANR-11-IDEX-0005-02). L.H.A.S. acknowledges partial funding from the research programme of the Netherlands eScience Center (www.nlesc.nl). The financial assistance of the South African SKA Project (SKA SA) towards this research is hereby acknowledged. Opinions expressed and conclusions arrived at are those of the authors and are not necessarily to be attributed to the SKA SA.

LOFAR, the Low Frequency Array designed and constructed by ASTRON, has facilities in several countries, that are owned by various parties (each with their own funding sources), and that are collectively operated by the International LOFAR Telescope (ILT) foundation under a joint scientific policy. 

The National Radio Astronomy Observatory is a facility of the National Science Foundation operated under cooperative agreement by Associated Universities, Inc.

\bsp

\label{lastpage}

\end{document}